\documentstyle[12pt, doublespace]{article}
\title{On the parameters of Lewis metric \\ \mbox{} \\ for the Lewis class}
\author{
\mbox{} \\
M. F. A. da Silva \\
{\footnotesize\em Departamento de F\'{\i}sica e Qu\'{\i}mica, 
Universidade Estadual Paulista,} \\
{\footnotesize\em Av. Ariberto Pereira da Cunha 333, 
12500 Guaratinguet\' a -- SP, Brazil and} \\
{\footnotesize\em Departamento de Astrof\'{\i}sica,
CNPq-Observat\'orio Nacional,} \\
{\footnotesize\em Rua General Jos\' e Cristino 77, 
20921-400, Rio de Janeiro -- RJ, Brazil.} \\
\mbox{} \\
L. Herrera \\
{\footnotesize\em Departamento de F\'{\i}sica, Facultad de Ciencias,} \\
{\footnotesize\em Universidade Central de Venezuela and} \\
{\footnotesize\em Centro de F\'{\i}sica, 
Instituto Venezolano de Investigaciones Cient\'{\i}ficas,} \\
{\footnotesize\em Caracas, Venezuela. 
Postal address: Apartado 80793, Caracas 1080A, Venezuela} \\
\mbox{} \\
F. M. Paiva and N. O. Santos \\
{\footnotesize\em Departamento de Astrof\'{\i}sica, 
CNPq-Observat\'orio Nacional,} \\
{\footnotesize\em Rua General Jos\'e Cristino 77, 
20921-400 Rio de Janeiro -- RJ, Brazil} \\
\mbox{} \\
{\footnotesize Author's internet addresses respectively: 
mfas@on.br,} \\ 
{\footnotesize lherrera@conicit.ve, fmpaiva@on.br 
and nos@on.br}}
\begin{document}
\setstretch{.8}
\maketitle
\date{}
\setstretch{1.6}

\newpage

\begin{abstract}
The physical and geometrical meaning of the four parameters of Lewis
metric for the Lewis class are investigated. Matching this spacetime to
a completely anisotropic, rigidly rotating, fluid cylinder, we obtain
from the junction conditions that the four parameters are
related to the vorticity of the source. Furthermore it is shown that
one of the parameters  must vanish if one wishes to reduce the
Lewis class to a locally static spacetime. Using the Cartan scalars it
is shown that the Lewis class does not include globally Minkowski as special
class, and that it is not locally equivalent to the Levi-Civita metric.
Also it is shown that, in contrast with the Weyl class, the parameter
responsible for the vorticity appears explicitly in the expression
for the Cartan scalars. Finally, to enhance our understanding  of the
Lewis class, we analyse the van Stockum metric.
\end{abstract}

\section{Introduction}  \setcounter{equation}{0}

In a recent paper \cite{SilvaHerreraPaivaSantos1994a}, we have
discussed  about the physical meaning of the four parameters of the
Lewis metric for the Weyl class. In this work we endeavour to extend
such discussion to the Lewis class. Such an effort will be justified,
in part, by the deep differences, exhibited below, between the two
classes.

Matching the Lewis class to a cylindrical fluid, a relationship linking
the vorticity of the source with three of the real constants entering into the
definition of the four complex parameters of the Lewis class
is found. Furthermore it will be shown that the vanishing of the
vorticity implies the vanishing of the parameter responsible for the
non-staticity of the metric.

From the study of the Cartan scalars it will be shown that, in
contrast with the Weyl class, the Lewis class is locally
distinguishable from the Levi-Civita metric. Also it will be shown the
Lewis class does not include locally flat spacetime as special
class. Therefore topological strings (in locally flat spacetime)
cannot be associated with this metric, hindering thereby the
topological interpretation of some of the parameters, in contrast with
the Weyl class \cite{SilvaHerreraPaivaSantos1994a}.

The three classes of van Stockum metric
\cite{vanStockum1937,Bonnor1980} provide an excellent example for our
discussion. We shall propose a lower limit for the linear mass density,
representing the frontier between the Lewis class and the Weyl class for
the van Stockum metric.

\section{Spacetime} \setcounter{equation}{0}

The spacetime is divided into two regions: the interior, with $0\leq r
\leq R$, to a cylindrical $\Sigma$ surface of radius $R$ centered along
$z$; and the exterior, with $R \leq r < \infty$. Both regions are
described by the general cylindrically symmetric stationary metric
\begin{equation}
ds^2=-f dt^2 +2kdtd\varphi +e^{\mu} \left( dr^2 +dz^2 \right)  
+ld\varphi^2,\label{2.1} 
\end{equation} 
where $f$, $k$, $\mu$ and $l$ are functions only of 
$r$, and the ranges of the coordinates $t$, $z$ and $\varphi$ are 
\begin{equation}
-\infty<t<\infty,\qquad -\infty<z<\infty,\qquad 0\leq \varphi \leq
2\pi,\label{2.2} 
\end{equation} 
with the hypersurfaces $\varphi=0$ and $\varphi=2\pi$ being identified.
The coordinates are numbered 
\begin{equation}
x^0=t,\qquad x^1=r,\qquad x^2=z, \qquad x^3=\varphi.\label{2.3} 
\end{equation}
Einstein's field equations 
\begin{equation}
R_{\mu\nu}=\kappa \left( T_{\mu\nu}-{1\over 2} g_{\mu\nu}T \right)\label{2.4} 
\end{equation} 
will be imposed to the metric (\ref{2.1}). The components of the Ricci
tensor $R_{\mu\nu}$ for (\ref{2.1}) are given in the reference
\cite{SilvaHerreraPaivaSantos1994a}. The exterior spacetime is
constituted of vacuum, hence Einstein's equations $(2.4)$ reduce to
\begin{equation}
R_{\mu\nu}=0.\label{2.5} 
\end{equation}
The general solution of (\ref{2.5}) for (\ref{2.1}) is the stationary
Lewis metric \cite{Lewis1932}, which can be written as \cite{KSMH1980} 
\begin{eqnarray}
f &=& ar^{-n+1} - {c^2\over{n^2 a}}r^{n+1}, \label{2.6} \\
k &=& -Af, \label{2.7} \\  
l &=&{r^2 \over f} -A^2 f,\label{2.8} \\ 
e^\mu &=& r^{{1\over 2}\left(n^2-1\right)}, \label{2.9}
\end{eqnarray}
with 
\begin{equation}
A={cr^{n+1}\over {naf}} + b. \label{2.11} 
\end{equation}
The constants $n$, $a$, $b$ and $c$ can be either real or complex, the
corresponding solutions belong to the Weyl class or Lewis class,
respectively. The Weyl class was studied in
\cite{SilvaHerreraPaivaSantos1994a}. Here we restrict our study to the
Lewis class. In this case, these constants are given by
\begin{eqnarray}
n &=& i\,m, \label{2.12} \\
c &=& {m\over 2}\left(a_1^2+b_1^2\right), \label{2.13} \\
a &=& {1\over 2}\left(a_1^2-b_1^2\right)+i\, a_1b_1, \label{2.14} \\
b &=& {a_1a_2+b_1b_2\over a_1^2+b_1^2}+{i\, \over a_1^2+b_1^2}, \label{2.15}
\end{eqnarray}
where $m$, $a_1$, $b_1$, $a_2$ and $b_2$ are real constants and satisfy
\begin{equation}
a_1b_2-a_2b_1=1.\label{2.10} 
\end{equation}
The equations ({\ref{2.12})--(\ref{2.10}) reveal us that if it is known
the value of the parameters $n$ and $a$, or $n$ and $b$, we can obtain
the parameter $c$. However knowing $n$ and $c$ we cannot obtain $a$ and
$b$.  The metric coeficients (\ref{2.6})--(\ref{2.9}) with
(\ref{2.12})--(\ref{2.15}) become \cite{Lewis1932}
\begin{eqnarray}
f &=& r\left(a_1^2-b_1^2\right)\cos{(m\ln{r})}+2r a_1 b_1\sin{(m\ln{r})},
\label{2.6a} \\
k &=& -r\left(a_1a_2-b_1b_2\right)\cos{(m\ln{r})}-r
\left(a_1b_2+a_2b_1\right)\sin{(m\ln{r})}, \hspace{1em}\label{2.7a} \\
l &=& -r\left(a_2^2-b_2^2\right)\cos{(m\ln{r})}-2r a_2b_2\sin{(m\ln{r})},
\label{2.8a} \\
e^\mu &=&r^{-{1\over2}\left(m^2+1\right)}. \label{2.9a} 
\end{eqnarray}
In fact, this metric is a subclass of the Kasner type metrics, as
pointed out in \cite{McIntosh1992}.

\section{Vorticity} \setcounter{equation}{0}

Using the transformation
\begin{equation}
d\varphi=d\bar\varphi+\omega dt,\quad {\rm where}\quad \omega=-{k\over l},
\label{2.16} 
\end{equation}
the metric (\ref{2.1}) can be diagonalized.  In order to have an
integral coordinate transformation $\omega$ must be constant, therefore
from equations (\ref{2.7a})--(\ref{2.8a}), $m=0$. This implies, from
(\ref{2.12})--(\ref{2.13}), that $n=0$ and $c=0$.  Thus the line
element becomes
\begin{equation}
ds^2 = {r\over a_2^2-b_2^2}dt^2 + r^{-{1\over 2}}\left(dr^2 + dz^2\right) -
r\left(a_2^2-b_2^2\right)d\bar\varphi^2 .\label{2.17} 
\end{equation}
This is a particular case of the static Levi-Civita metric with the
energy density per unit length $\sigma$, given by (\ref{5.26}),
equals~${1\over4}$. Nevertheless the transformation (\ref{2.16}) is not
global, since the new coordinate $\bar\varphi$ ranges from $-\infty$ to
$\infty$ instead of ranging from $0$ to $2\pi$
\cite{SilvaHerreraPaivaSantos1994a,Stachel}.

Considering the interior spacetime of the cylinder, $0\leq r\leq R$,
filled with anisotropic fluid, then we can integrate one of the
Einstein's equations (\ref{2.4}) and obtain
\begin{equation}
\xi r =fk^\prime - kf^\prime,\label{2.18} 
\end{equation}
where $\xi$ is a constant. $\xi$ measures the vorticity of the source,
since a straightforward calculation
\cite{SilvaHerreraPaivaSantos1994a} shows that the magnitude of the
vorticity tensor is $\xi/(2fe^{{\mu\over 2}})$. Considering $f$ and $k$
given by (\ref{2.6a}) and (\ref{2.7a}), we have
\begin{equation}
\xi = -m\left(a_1^2+b_1^2\right).\label{2.19} 
\end{equation}
So,
\begin{equation}
c=-{\xi\over 2}.\label{2.20}
\end{equation}
Hence in order to have the vorticity equals to zero, i.e. $\xi=0$, 
we need $m=0$ since $a_1^2+b_1^2\neq 0$.

Observe the difference, at this point, between the Weyl class
\cite{SilvaHerreraPaivaSantos1994a} and the Lewis class. In the latter
the vanishing of the vorticity yields a locally Levi-Civita spacetime,
whereas in the former the vanishing of vorticity does not,
necessarily, implies that the metric can be reduced to a globally or
locally static spacetime.

\section{The Cartan scalars}  \setcounter{equation}{0}

It is known \cite{MacCallumSkea1993} that the so called 14 algebraic
invariants (and even all the polinomial invariants of any order) are
not sufficient for locally characterizing a spacetime, in the sense
that two metrics may have the same set of invariants and be not
equivalent. As an example, all these invariants vanish for both
Minkowski and plane-wave \cite{Schmidt1994,MacCallumSkea1993}
spacetimes and they are not the same. A complete local
characterization of spacetimes may be done by the Cartan
scalars. Briefly, the Cartan scalars are the components of the Riemann
tensor and its covariant derivatives (up to possibly the $10^{\rm th}$
order) calculated in a constant frame. For a review, see
\cite{SilvaHerreraPaivaSantos1994a} and references therein. In
practice, the Cartan scalars are calculated using the spinorial
formalism. For the purpose here, the relevant quantities are the Weyl
spinor $\Psi_A,$ and its first covariant symmetrized derivative
$\nabla\Psi_{AB'}$, which represent the Weyl tensor and its covariant
derivative. It should be stressed that, although the Cartan scalars
provide a local characterization of the spacetime, global properties
such as topological deffects do not probably appear in them.

In a previous paper \cite{SilvaHerreraPaivaSantos1994a} the Cartan
scalars for the Weyl class of Lewis metric are given. The Lewis
class metric may be obtained from the Weyl class metric by considering the
constants $a$, $b$, $c$ and $n$ as complex, subjected to side relations
(\ref{2.12})--(\ref{2.10}) to assure that the metric components remain
real. Therefore, the Cartan scalars for the Lewis class can be obtained
from those of the Weyl class by a proper redefinition of the constants.
As in the Weyl class, only the constant $n$ appears in the Cartan
scalars. Nevertheless, here, $n$ must be substituted by its complex
value (\ref{2.12}) $\,i\,m$. The nonvanishing Cartan scalars are:
 \begin{equation} \label{CS}
\begin{array}{rcl}
\Psi_2 &=& {1\over 8}(m^2+1)r^{{1\over 2}(m^2-3)}, \\
\Psi_0 =\Psi_4 &=& -i\,m\Psi_2, \\ 
\nabla\Psi_{01'} =\nabla\Psi_{50'} &=& 
-{\sqrt {2}\over {16}}i\,m(m^4-1)r^{{3\over 4}(m^2-3)}, \\
\nabla\Psi_{10'}=\nabla\Psi_{41'}&=&
{\sqrt {2}\over 8}i\,m(m^2+1)r^{{3\over 4}(m^2-3)}, \\
\nabla\Psi_{21'}=\nabla\Psi_{30'} &=& 
{\sqrt {2}\over {32}}(m^2-3)(m^2+1)r^{{3\over 4}(m^2-3)}. 
\end{array}
\end{equation}
This provides an invariant criterion distinguishing the Lewis and the
Weyl classes, since for the Lewis class, for instance,
$\Psi_0=-im\Psi_2$ while, for the Weyl class
\cite{SilvaHerreraPaivaSantos1994a}, $\Psi_0=-n\Psi_2$, where $m$ and
$n$ are arbitrary real constants.

Contrary to the Weyl class [1], the Cartan scalars for the Lewis class
are distinguishable from those of the Levi Civita metric, except for
$m=0$ (cf. section 3). Furthermore there is no value of $m$ for which
the Cartan scalars are all zero, implying at once that the Lewis class
does not include Minkowski as special class.

On the other hand, the van Stockum exterior solution
\cite{vanStockum1937,Bonnor1980} (case I), which is a particular case
of the Lewis metric, contains the globally Minkowski spacetime as
special case.\footnote {Although in the form presented in this paper,
it is not obvious that this special case occurs, it is quite simple to
obtain this limit in the original form \cite{Bonnor1980}.}  Therefore
the van Stockum solution (case I) must be a particular case of the
Weyl class. Since the case I of the van Stockum exterior solution
cannot be reduced to the globally static Levi-Civita metric
\cite{Bonnor1980} (neither case II and III), it is clear that it is a
particular case of the Weyl class with $b\neq 0$ and $c\neq 0$, since
for $b=0$ and $c=0$ the Weyl class can be globally reduced to the
static Levi-Civita metric. In the next section we shall make a more
detailed analysis of the van Stockum solution, which will help us to
improve our comprehension on the parameters of the Lewis metric.

\section{The van Stockum's metric classification}  \setcounter{equation}{0}

In 1937 van Stockum \cite{vanStockum1937} solved the problem of a
rigidly rotating infinite cylinder filled with dust and matched it to
the vacuum Lewis solution. The solution depends on the parameter $wR$,
related to the mass per unit length of the dust cylinder (note that in
\cite{vanStockum1937,Bonnor1980}, the letter $a$ is used instead of
$w$), and is given by, for $wR<{1\over 2}$ (case I),
\begin{eqnarray}
f &=& -r\left[
2\beta \cosh{(2N \ln{r})}+{\alpha^2+\beta^2\over\alpha}\sinh{(2N \ln{r})}
\right],\label{5.1}\\
k &=& -r\left[
\cosh{(2N \ln{r})}+{\beta\over \alpha}\sinh{(2N \ln{r})}\right],\label{5.2}\\
l &=& {r\over\alpha} \sinh{(2N \ln{r})},\label{5.3}\\
e^\mu &=& 
\lambda\left({r\over R}\right)^{\left(2N^2-{1\over 2}\right)},\label{5.4}
\end{eqnarray}
and for $wR>{1\over 2}$ (case III),
\begin{eqnarray}
f &=& r\left[
2\beta \sin{(2N \ln{r})}+{\alpha^2-\beta^2\over \alpha}\cos{(2N \ln{r})}
\right],\label{5.5}\\
k &=& r\left[
\sin{(2N \ln{r})}-{\beta\over\alpha}\cos{(2N \ln{r})}\right],\label{5.6}\\
l &=& {r\over \alpha}\cos{(2N \ln{r})},\label{5.7}\\
e^\mu &=& 
\lambda\left({r\over R}\right)^{-\left(2N^2+{1\over 2}\right)},
\label{5.8}
\end{eqnarray}
The constants $\alpha$, $\beta$, $N$ and $\lambda$ are given by, for case I,
\begin{eqnarray}
N &=& {1\over 2}\sqrt{1-4w^2 R^2},\label{5.9}\\
\alpha &=& {\sqrt{1-4w^2R^2}\over 2w^3R^4},\label{5.10}\\
\beta &=& -{1-2w^2R^2\over 2w^3R^4},\label{5.11}\\
\lambda &=& e^{-w^2 R^2},\label{5.12}
\end{eqnarray}
and for case III,
\begin{eqnarray}
N &=& {1\over 2}\sqrt{4w^2 R^2-1},\label{5.13}\\
\alpha &=& {\sqrt{4w^2 R^2-1}\over 2w^3 R^4},\label{5.14}\\
\beta &=& {2w^2 R^2-1\over 2w^3 R^4},\label{5.15}\\
\lambda &=& e^{-w^2 R^2}.\label{5.16}
\end{eqnarray}
Case II, i. e., $wR={1\over 2}$ is defined by van Stockum
\cite{vanStockum1937} by a limiting process of case I. We add that it
is also a limit of case III. Nevertheless, it should be stressed that
the direct subtitution of  $wR={1\over 2}$ in cases I or III does not
give the proper result.  The van Stockum solution I belongs to the Weyl
class, where the real parameters $n$, $a$, $b$ and $c$ assume the
following values,
\begin{eqnarray}
n&=&\sqrt{1-4w^2R^2},\label{5.17}\\
a&=&{(\alpha-\beta)^2\over 2\alpha},\label{5.18}\\
b&=&\pm{1\over \alpha-\beta},\label{5.19}\\
c&=&{(\alpha^2-\beta^2)\over\alpha}N, \label{5.20}
\end{eqnarray}
while the van Stockum solution III belongs
to the Lewis class, where the real parameters $m$, $a_1$, $b_1$, $a_2$
and $b_2$ assume the following values,
\begin{eqnarray}
m &=& \sqrt{4w^2R^2-1},\label{5.21}\\
b_2 &=& 0,\label{5.22}\\
a_2 &=& -{1\over b_1},\label{5.23}\\
a_1 &=& {\beta\over b_1},\label{5.24}\\
b_1^2 &=& -\alpha.\label{5.25}
\end{eqnarray}

In order to understand better the Lewis class metric let us consider
shortly the Weyl class metric. In \cite{SilvaHerreraPaivaSantos1994a}
it is shown that the Newtonian mass per unit length is given by
\begin{equation}
\sigma={1\over 4}(1-n),\label{5.26}
\end{equation}
with $n$ being a real constant.
So, from (\ref{5.17}) and (\ref{5.26}), we have
\begin{equation}
\sigma={1\over 4}\left[1-\sqrt{1-4w^2R^2}\right] \label{5.27}
\end{equation}
for case I. For $w^2R^2\ll1$, this expression reduces to 
\begin{equation}
\sigma={1\over 2}w^2R^2.\label{5.28}
\end{equation}
This is the same value obtained by Bonnor \cite{Bonnor1980} in this
aproximation. Using this result, he establishes a lower limit for the
linear mass density in case III ($wR>{1\over 2}$), obtaining ${1\over
8}$. We believe that a better lower limit would be given directly by
(\ref{5.27}), which is $\sigma={1\over 4}$. Then $\sigma={1\over 4}$
represents the frontier between the Weyl class metric and the Lewis
class metric at least for the particular case of van Stockum.

Returning to the Lewis class metric, it is important to note that the
Cartan scalars do not admit Minkowski spacetime. This is in accordance
with the existence of a lower limit for $\sigma$ in the van Stockum
solution III, since with this lower limit the source cannot be made
vaccum and therefore the exterior solution cannot be Minkowski.

The Cartan scalars impose a superior limit to the parameter $m$, given by
\begin{equation}
m\leq \sqrt{3},\label{5.30}
\end{equation}
since for $m$ larger than this value, the singularity is at
 $r=\infty$, not in $r=0$.  When we substitute this value in
 (\ref{5.21}), considering the equality, we have that $wR=1$, which
 agrees with Bonnor's result \cite{Bonnor1980}.

\section{Conclusion}  \setcounter{equation}{0}

It is known that the Lewis metric comprises two differents families
called Weyl class and Lewis class. The first one occurs when we
consider that all four parameters appearing in the metric are real. On
the other hand, in the second one, some of the parameters may be
complex. On figure 1 we present a diagram showing some of the
subclasses of the Lewis metric, according to the value of the
parameters.

In a previous paper \cite{SilvaHerreraPaivaSantos1994a} we obtained
physical interpretations for the four real parameters $n$, $a$, $b$
and $c$ which characterize the Lewis metric for the Weyl class. The
parameters $b$ and $c$ are related to the non-staticity of the
spacetime, since when we take $b=0$ and $c=0$ the Weyl class reduces
to the static Levi-Civita metric. While the parameter $n$ is
associated to the Newtonian mass per unit length of an uniform line
mass $\sigma$ when it produces low densities. So if we assume $n=1$,
which means $\sigma=0$, the Weyl class metric becomes a static locally
flat spacetime or a stationary locally flat spacetime if we take $b=0$
and $c=0$ or $b\neq 0$ and $c\neq 0$, respectively. The parameter $c$
measures the vorticity of the source of the Weyl class metric when it
is matched to a general stationary completely anisotropic fluid.
Finally, the parameter $a$ was interpreted as the constant arbitrary
potential that exists in the corresponding Newtonian solution and a
linear energy density along a string in the locally flat static limit
of the Weyl class. When $a=1$ the locally flat static spacetime
reduces to the globally Minkowski spacetime.

In this paper we give more attention to the Lewis class metric. We
show that, as in the Weyl class, the parameter $c$ is also
proportional to the vorticity of the source of the Lewis class metric
if we match this spacetime to the same general stationary completely
anisotropic fluid. Moreover for the Lewis class we find that the
vorticity is also directly associated with the parameter $n$ and
indirectly related with the parameters $a$ and $b$ and, in fact, the
vorticity vanishes if $m=0$. We verify that the van Stockum solution
for vacuum is a particular case of the Lewis metric. Indeed, the van
Stockum case I is a subclass of the Weyl class metric while the van
Stockum case III is a subclass of the Lewis class metric. The case II
of the van Stockum solution can be obtained from the case I or from
the case III by a limiting process when $n\rightarrow 0$.  As can be
seen in the Cartan scalars, the Lewis class can never be reduced to
the locally flat spacetime.

\section{Acknowledgment}
MFAS and FMP gratefully acknowledge financial assistance from CAPES
and CNPq, respectively.

\begin{figure}
\setlength{\unitlength}{3ex}
%\begin{picture}(26,19)
%\begin{picture}(19,13.5)(-0.8,0)
\begin{picture}(19,15)(-1.6,0)

\put(9,18){\framebox(5,2)[t]}
\put(9,17.9){\makebox(5,2)[t]{{\bf Lewis metric}}}
\put(9,17){\makebox(5,2)[t]{$n$, $a$, $b$, $c$}}

\put(1,15){\framebox(5,2)[t]}
\put(1,14.9){\makebox(5,2)[t]{Weyl class}}
\put(17,15){\framebox(5,2)[t]}
\put(17,14.9){\makebox(5,2)[t]{Lewis class}} 
\put(1,14){\makebox(5,2)[t]{real $n$}}
\put(17,14){\makebox(5,2)[t]{imaginary $n$}}
\put(11,17.8){\vector(-3,-1){4.7}}
\put(12,17.8){\vector(3,-1){4.8}}

\put(3,6){\framebox(4.5,2)[t]}
\put(3,5.9){\makebox(4.5,2)[t]{Levi-Civita}}
\put(3,5){\makebox(4.5,2)[t]{$b=0$, $c=0$}}
\put(4.8,14.8){\vector(0,-1){6.5}}

\put(6,12){\framebox(5.4,2)[t]}
\put(6,11.9){\makebox(5.4,2)[t]{van Stockum I}}
\put(6,11){\makebox(5.4,2)[t]{$\ast$}}
\put(5.5,14.8){\vector(3,-2){0.8}}

\put(13,12){\framebox(5.6,2)[t]}
\put(13,11.9){\makebox(5.6,2)[t]{van Stockum III}}
\put(13,11){\makebox(4.6,2)[t]{$\ast\ast$}}
\put(19,14.8){\vector(-3,-2){0.8}}

\put(15,6){\framebox(7,2)[t]}
\put(15,5.9){\makebox(7,2)[t]{locally Levi-Civita}}
\put(15,5){\makebox(7,2)[t]{$n=0$}}
\put(19.5,14.8){\vector(0,-1){6.5}}
%\put(7.8,7){\vector(1,0){7,0}}

\put(9.5,9){\framebox(6.5,2)[t]}
\put(9.5,8.9){\makebox(6.5,2)[t]{van Stockum II}}
\put(9.5,8){\makebox(6.5,2)[t]{$n\rightarrow 0$ and $\ast$ or $\ast\ast$}}
\put(10,11.8){\vector(4,-3){0.8}}
\put(15,11.8){\vector(-4,-3){0.8}}

\put(-1,12){\framebox(5.4,2)[t]} 
\put(-1,11.9){\makebox(5.4,2)[t]{Flat stationary}} 
\put(-1,11){\makebox(5.4,2)[t]{$n=\pm 1$}}
\put(2.5,14.8){\vector(-3,-2){0.8}}

\put(1,3){\framebox(7,2)[t]}
\put(1,2.9){\makebox(7,2)[t]{Flat static}} 
\put(1,2){\makebox(7,2)[t]{$n=\pm 1$, $b=0$, $c=0$}}
%\put(1.1,14.8){\vector(0,-1){10}}
\put(2,11.8){\vector(0,-1){6.5}}
\put(4.8,5.8){\vector(0,-1){0.6}}

\put(3,0){\framebox(10,2)[t]}
\put(3,-0.1){\makebox(10,2)[t]{globally Minkowski}}
\put(3,-1){\makebox(10,2)[t]{$n=\pm 1$, $b=0$, $c=0$, $a=1$}}
\put(8.5,11.8){\vector(0,-1){9.5}}
\put(4.8,2.8){\vector(0,-1){0.6}}

\end{picture}
\caption[]{Lewis metrics and their subclasses. Note that flat means
locally flat. $\ast$ means that equations (\ref{5.17})--(\ref{5.20})
should be satisfied. $\ast\ast$ means that equations
(\ref{5.21})--(\ref{5.25}) should be satisfied.}
\label{fig} 
\end{figure}
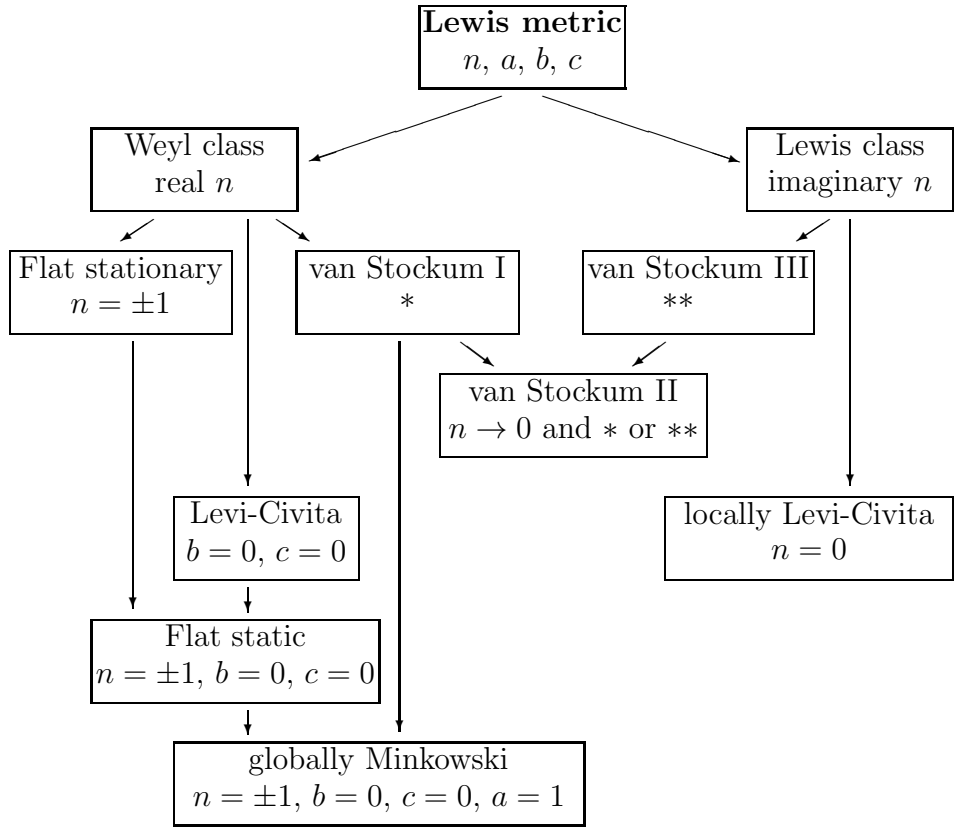

\appendix

\section{Appendix} \setcounter{equation}{0}

We present below the components of the Riemann tensor for the Lewis
class, in the same Lorentz frame used in
\cite{SilvaHerreraPaivaSantos1994a}
\begin{equation} \label{Rie}
\begin{array}{rcl}
R_{0101}=-R_{2323} &=& {1\over 4}(m^2+1)r^{{1\over 2}(m^2-3)} \\
R_{1313}=-R_{0202} &=& 
{1\over 8}(i\,m+1)(m^2+1)r^{{1\over 2}(m^2-3)} \\
R_{0303}=-R_{1212} &=& 
{1\over 8}(i\,m-1)(m^2+1)r^{{1\over 2}(m^2-3)}
\end{array}
\end{equation}
Note that some of these components are complex. This present no problem
since the frame used becomes complex for the Lewis class. For
completeness, we list the nonvanishing algebraic invariants, which are
of course real:  
\begin{equation} \label{14}
\begin{array}{rcl}
R_{\alpha\beta\gamma\delta}R^{\alpha\beta\gamma\delta} &=& 
-{1\over 4}(m^2-3)(m^2 +1)^2 r^{(m^2-3)} \\
R_{\alpha\beta\gamma\delta}R^{\gamma\delta\mu\nu}{R_{\mu\nu}}^{\alpha\beta}
&=&
{3\over 16}\left(m^2+1\right)^4 r^{{3\over 2}(m^2-3)}
\end{array}
\end{equation}

\end{document}